\documentclass[preprints,article,accept,moreauthors,pdftex]{Definitions/mdpi}

\firstpage{1} 
\pubvolume{xx}
\issuenum{1}
\articlenumber{5}
\pubyear{2019}
\copyrightyear{2019}
\newcommand{\modified}[1]{\textcolor{black}{#1}}
\history{}

\pdfoutput=1

\usepackage{circuitikz}
\Title{Novel Approach To Synchronisation Of Wearable IMUs 
Based On Magnetometers}

\Author{Andreas Spilz $^{1}$\orcidA{} and Michael Munz $^{2,*}$\orcidB}

\AuthorNames{Andreas Spilz and Michael Munz}

\address{%
$^{1}$ \quad Laboratory for Biomechatronics, University of Applied Sciences Ulm, 89081, Germany; Andreas.Spilz@thu.de\\
$^{2}$ \quad Laboratory for Biomechatronics, University of Applied Sciences Ulm, 89081, Germany; Michael.Munz@thu.de}

\corres{Correspondence: Michael.Munz@thu.de}

\abstract{Synchronisation of wireless inertial measurement units in human movement analysis is often achieved using event-based synchronisation techniques. However, these techniques lack precise event generation and accuracy. An inaccurate synchronisation could lead to large errors in motion estimation and reconstruction and therefore wrong analysis outputs. We propose a novel event-based synchronisation technique based on a magnetic field, which allows sub-sample accuracy. A setup featuring Shimmer3 inertial measurement units is designed to test the approach. The proposed technique shows to be able to synchronise with a maximum offset of below 2.6~ms with sensors measuring at 100~Hz. Also, the results indicate a reliable event generation and detection. The investigated parameters suggest a required synchronisation time of eight seconds. Further research \modified{should investigate the temperature changes that the sensors are exposed to during human motion analysis and their influence on the internal time measurement of the sensors. In addition, the approach should be tested using inertial measurement units from different manufacturers} to investigate an identified constant offset in the accuracy measurements.}


\begin{document}

\section{Introduction}
\label{section:1}

Today, wireless inertial measurement units (IMU) are often used for human motion analysis tasks like gait analysis \cite{Weygers.2020, VienneJumeau.2020, Poitras.2019, OReilly.2018, Ghislieri.2019}. \modified{In such analysis, a large number of wireless IMUs are used. For example XSens Technologies (Enschede, Netherlands), a well-known manufacturer of IMU-based whole-body tracking solutions, uses setups with up to 17 IMUs. Motion analysis tasks are usually conducted over a period of a few minutes to several hours. Therefore,} an important question to answer when designing an appropriate experiment, is how the timestamps of the wireless sensors are synchronised. \modified{The timestamps of an individual unit are determined on the basis of an internal oscillator. The frequency of these oscillators varies due to diverse influences, such as the ambient temperature, manufacturing tolerances or aging. Those influences sum up to a drift, which therefore contains linear and non-linear components. Over time, this can lead to large temporal deviations between the sensor units \cite{Tirado-Andres.2019}.} If the timestamps of the sensors are not synchronous, and the fused sensor data is taken into account for the motion analysis, large processing errors can occur.

Time synchronisation in wireless sensor networks can be achieved with established network protocols. Examples for prominent protocols are the Reference Broadcast Scheme \cite{Elson.2002} (RBS), the Timing-sync Protocol for Sensor Networks \cite{Ganeriwal.2003} (TPSN) \modified{and the Flooding Time
Synchronisation Protocol (FTSP) \cite{FTSP.2004}.} However, these protocols are dependent on network capabilities of the used sensors, which are not available in most wireless IMUs focusing on applications for human motion analysis \cite{Zhou.2020}. These protocols also cause other issues like additional computational cost, resulting in increased battery consumption \cite{Sivrikaya.2004b}. The Shimmer3 IMUs from ShimmerSensing \modified{(Dublin, Ireland)} used in this work have an integrated Bluetooth-based synchronisation protocol. However, if Shimmer's built in synchronisation is enabled, the data streaming features are disabled. Other Bluetooth-based protocols like \cite{Ringwald.2007} require access to the Bluetooth module's clock, which is not granted by most device firmwares.
To avoid these issues, it is common practice to use event-based synchronisation techniques \cite{Bannach.2009}. The individual sensor nodes are synchronised by generating and detecting events in the accelerometer and gyrometer data \cite{Gao.2019, Rietveld.2019, Paraschiakos.2020, HarryJ.Witchel.2018, Kim.2013, Chen.2013}. For example, the subject wearing the IMUs performs a jump and the resulting spikes in acceleration are used to synchronise the sensors. This method is time efficient and easy to implement, yet there are disadvantages: It is difficult to generate precise physical impulses which result in an unambiguously interpretable event in the data. The main drawback is that the methods lack accuracy, because the used sample rates in human movement analysis are generally around 100~Hz \cite{Zhou.2020}, which results in a possible synchronisation uncertainty of 10~ms. Hence there is a need for an event-based synchronisation technique for IMUs which relies on accurate detectable physical impulses and is able to synchronise the sensor nodes with sub-sample accuracy. 
The purpose of this article is to present and evaluate an appropriate novel approach, which uses a magnetic field \modified{and is suitable for human movement analysis tasks}. We investigate the accuracy of the presented approach and evaluate the influence of different key parameters on the accuracy. This research may provide an alternative to the commonly used event-based synchronisation techniques.

\modified{The structure of this paper is as follows: Section \ref{section:2} describes the theory behind the novel approach first, followed by the practical implementation of the setup and the design of the experiments to analyse the approach. Section \ref{section:3} presents the results of the validation experiments. Section \ref{section:4} discusses the experiments performed and derives the application area and limitations of the approach. Finally, section \ref{section:5} summarizes the key points of this paper.}
\section{Materials and Methods}
\label{section:2}
To address the mentioned issues with current event-based time synchronisation procedures we developed a novel synchronisation procedure. This procedure is based on the generation of a precise magnetic signal and a detection algorithm. The magnetic signal is measured simultaneously by each IMU to be synchronised and stored together with the IMU's timestamp. The signal's trace is designed in a manner that allows us to determine the beginning of the synchronisation procedure $t_0$ with sub-sample accuracy based on the recorded IMU data.

\begin{figure}[h]
    \centering
    \includegraphics[width=13 cm]{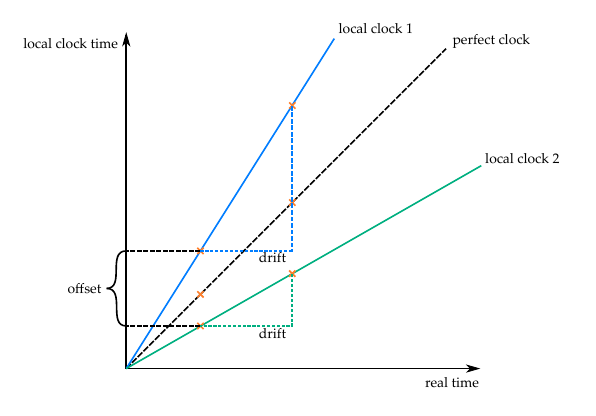}
    \caption{Compensating offset and linear \modified{component of the} drift in clocks. The orange x marks represent the beginning of the first and second synchronisation procedure in the respective time contexts. Using the local timestamps of the first synchronisation procedure it is possible to compensate the offset between the two exemplary clocks. With the timestamps of both synchronisation procedures one can compensate the linear \modified{component of the} clock drift.}
    \label{fig:explanation_drift}
\end{figure}

The described process is performed twice at the start and at the end of each measurement. After the measurement has been completed, the magnetic synchronisation signal is detected in the recorded data. An algorithm extracts the beginning of the two synchronisation procedures for each IMU in the local time context. Between these two points, the captured data can now be synchronised, which means compensating the offset and \modified{the linear component of the drift} of an IMU's local clock. In figure~\ref{fig:explanation_drift}, these two kinds of error are displayed, together with the events generated by the two synchronisation procedures. As can be seen in the figure, the offset between the local clocks of the IMUs can be compensated using the event generated by the first synchronisation procedure. Using both events it is also possible to compensate the relative linear drift between the IMUs. Compensating this \modified{linear component of the drift} is important for long-term measurements. To achieve this, we first compensate the time offset of each IMU to generate a common start point. Afterwards we map the IMU's timestamps onto the timestamps of one arbitrary chosen reference IMU. 

\subsection{Synchronisation Algorithm}
In the following, the signal generation process will be described in detail as well as the process that allows us to achieve sub-sample accuracy. When an inductor is connected or disconnected from a direct power source, the amount of current flowing changes in an asymptotic fashion. This behaviour is called the transient response of the inductor, and is described by the following equation:

\begin{equation}
    i(t) = I(1-e^{-\frac{R}{L} t} )
    \label{eq:1}
\end{equation}
L and R describe the inductance and the electrical resistance of the inductor. I denotes the maximum current delta of this inductor. It is calculated from the difference of the current flowing through the inductor between the connected and disconnected state.
The time constant $\tau$ is used as the time unit of the transient response. It can be calculated using the electrical properties of the inductor:

\begin{equation}
    \tau = \frac{L}{R}
    \label{eq:2}
\end{equation}

After about $5\tau$, $i(t)$ reaches a steady value. When the amount of current flowing through an inductor is changing, the measurable magnetic flux density $B$ in- or decreases proportionally. This relation to $i(t)$ is described by the following equation, which calculates $B$ based on the constant properties of the used inductor and the changing amount of current $i(t)$:

\begin{equation} 
    B(t)=\mu*\frac{N*i(t)}{l}
    \label{eq:3}
\end{equation}

N refers to the number of windings, l to the length of the used inductor and $\mu$ to the magnetic field constant. To measure $B$, the magnetometer of an IMU can be used. However, many magnetometers in commercially available IMUs are limited to a maximum sampling rate of 100~Hz. This limitation complicates capturing the mentioned behaviour of $B$, because most commercially available inductors have characteristics which result in a time constant $\tau$ significantly lower than 1 ms. Accordingly, the transient response takes up to about 5~ms and can only be captured irregularly at a sampling rate of 100~Hz.

If we can acquire a sample during the transient response, we would be able to calculate the beginning of the response using equation \ref{eq:3}. To achieve this, a circuit of parallel connected inductors was designed. Later, each IMU is placed above one inductor. This circuit uses a constant power source which can be disconnected with a transistor. Now the power source is periodically connected and disconnected with a low frequency. 
The frequency $f$ is selected so that the inductors are able to go through the whole transient response, resulting in: 

\begin{equation}
    f < \frac{1}{5\tau}
\end{equation}

In order to capture this transient response with the magnetometers it is important to select a frequency which is not a multiple of 100~Hz. These differing frequencies ensure that the IMUs eventually sample during the transient response. Samples which are acquired during a transient response will be referred to as $hits$ $h(i)$ in this paper, where $i$ represents the number of already performed transient responses. The timestamp and measured magnetic flux density of a hit are denoted by $t(i)$ and $k(i)$.

\begin{figure}[hb]
    \centering
    \includegraphics[width=15 cm]{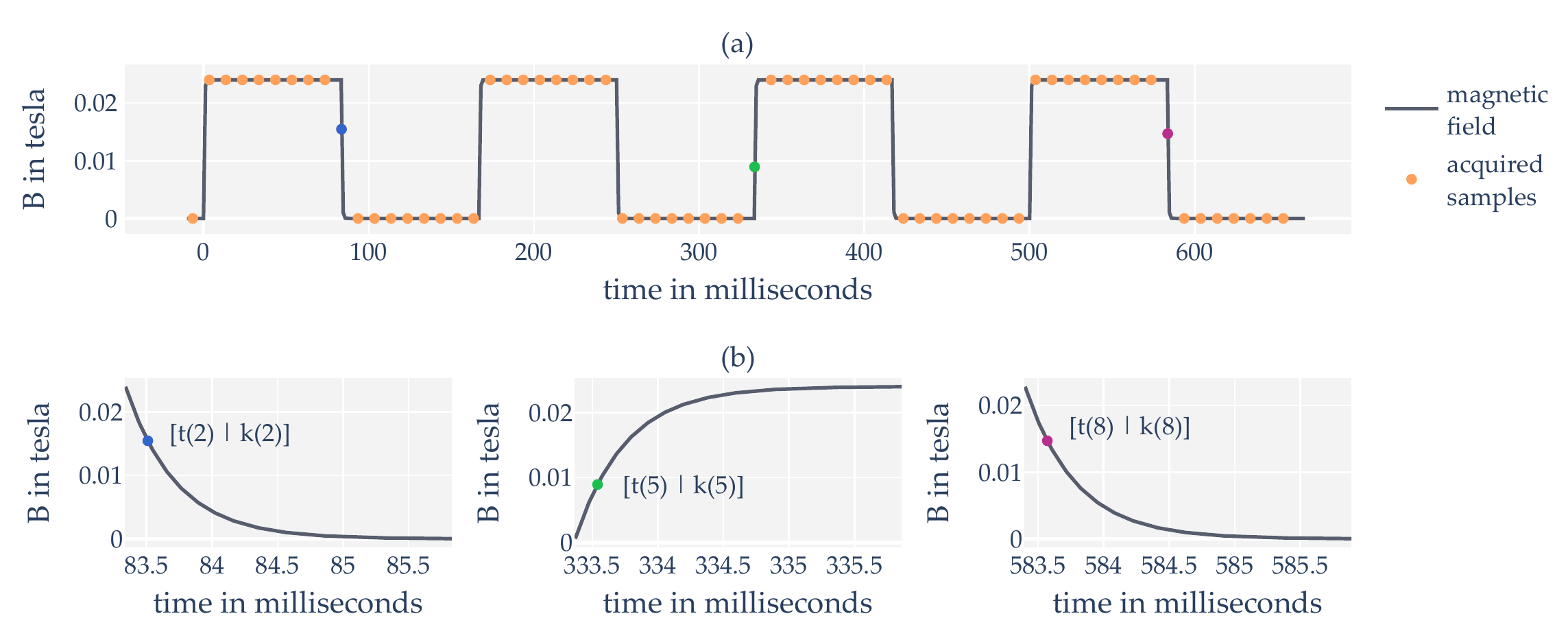}
    \caption{ Magnetic signal and resulting magnetometer data used to perform the synchronisation procedure. (a) Magnetic flux density $B$ generated by a transistor which is periodically (6~Hz) dis-/connected from a power source. Additionally, the samples acquired by a magnetometer measuring with 100~Hz are displayed. (b) Detail view of the three samples ($hits$) from the data in (a) which were acquired during the transient response.}
    \label{fig:alg1}
\end{figure}

\begin{figure}[h]
    \centering
    \includegraphics[width=15 cm]{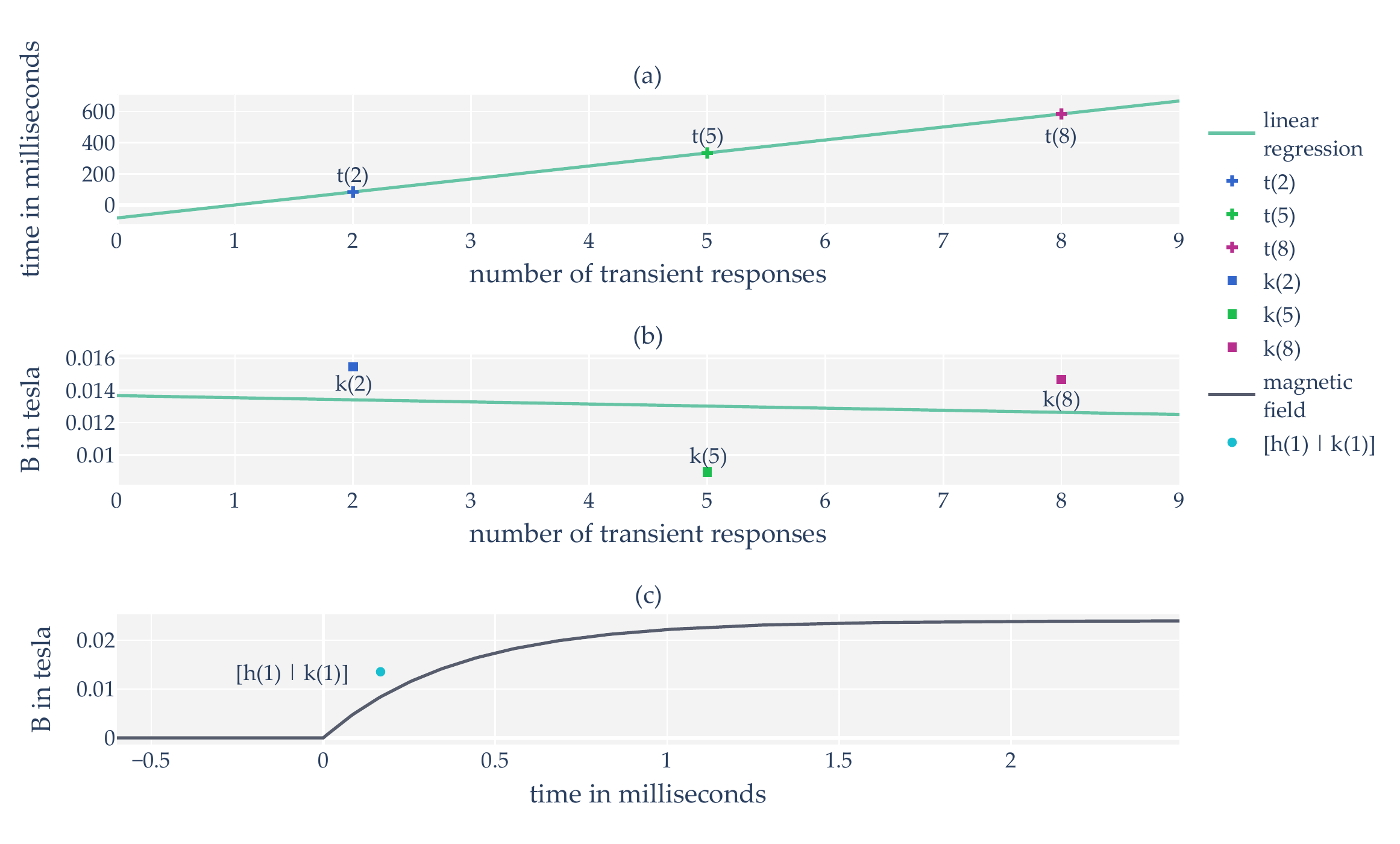}
    \caption{ Magnetic signal and resulting magnetometer data used to perform the synchronisation procedure. (a) The resulting function of a linear fit with the timestamps $t(i)$ of the hits and the number of already performed transient responses $i$. This function allows us to estimate $t(1)$. (b) The resulting function of a linear fit with the magnetic flux density of the hits $k(i)$ and the number of already performed transient responses $i$. This function allows us to estimate $k(1)$. (c) The estimated hit on the first transient response. There can be a small deviation between the calculated and the actual magnetic flux density.}
    \label{fig:alg2}
\end{figure}

This idea is illustrated in figure~\ref{fig:alg1}~(a), which shows the generated magnetic flux density and the acquired samples of an IMU. In this illustration there are three $hits$ which are displayed with a higher temporal resolution in figure~\ref{fig:alg1}~(b).
Using these $hits$ we now estimate the timestamp $t_0$ of the beginning of a synchronisation procedure. The beginning is marked by the start of the first transient response. First we compute the timestamp $t(1)$ of a $hit$ on this transient response. To achieve this we perform a polynomial fitting of first order with the timestamps of the individual hits $t(i)$ (see figure~\ref{fig:alg2}~(a)):

\begin{align}
    \boldsymbol{\theta} &= (\theta_1, \theta_2) \\
    \boldsymbol{\hat{\theta}} &= argmin_{\theta}(\sum[y_i-f(x_i)]^2) = argmin_{\theta}(\sum[y_i-(\theta_1 \cdot x_i+\theta_2)]^2) 
\end{align}

Under the assumption of a constant sampling interval, the resulting equation allows us to calculate the timestamp of a hit on any transient response. Due to clock drift in real sensors, this linear fit only provides an estimate for the timestamp for each hit $h(i)$. As the properties of the inductor are known, it is possible to calculate the time since the beginning of a transient response given a magnetic flux density. Therefore, we perform a second linear fit (see figure~\ref{fig:alg2} (b)) using the measured magnetic flux densities $k(i)$ and the number of already performed transient responses $i$ to estimate $k(1)$. In order to calculate the start time of the transient response with sub-sample accuracy, we substitute equation \ref{eq:1} in equation \ref{eq:3} and solve for $t_{TR}$ for a given magnetic flux density.
As $I$ cannot be measured by the IMUs, it is replaced by $K$, which is proportional to I. $K$ denotes the measurable difference in magnetic flux density before and after a transient response. These steps result in the following equation:

\begin{equation}
    \ t_{TR} =  \dfrac{-\ln \left( 1 -\frac{k(1) \cdot l}{\mu \cdot N \cdot K}  \right) \cdot L}{R}
\end{equation}

Finally, subtracting this result from $t(1)$ leads to the start of the synchronisation procedure with sub-sample accuracy:

\begin{equation}
   \ t_{0} = t(1) - t_{TR}
\end{equation}

In summary, the well-defined asymptotic behaviour of B during an inductor's transient response allows us to determine the start of a synchronisation procedure $t_0$ with sub-sample accuracy. As described at the beginning of this chapter, this synchronisation process is executed before and after each measurement. Two synchronous time points can thus be calculated for each IMU, on the basis of which the measurement data is synchronised.

\subsection{Experimental setup}
\label{section:tech}
\begin{figure}[h]
    \centering
     \begin{circuitikz}[european]
    \draw (0,3) node[npn ](Q){};
    \draw (Q.C) to [short, -] (0,4);
    \draw (Q.B) to [short, -o] ++ (-0.2, 0) node[left](vi1){control signal};
    \draw (0,4) to [short, -] (7.5, 4);
    \draw [dashed] (7.5, 4) -- ++ (1.5, 0) coordinate(d);
    \draw (d) to [short, -] ++(2.5, 0);
    \draw
     (0,0) to [short, -] ++(3,0)
  to [L, l=$L_1$] ++(0,4)
  (Q.E) to [V, l=$5~V$] (0,0)
  (3,0) to [short, -] ++(2,0)
  to [L, l=$L_2$] ++(0,4)
  (5,0) to [short, -] ++(2,0)
  to [L, l=$L_3$] ++(0,4)
  (7,0) to [short, -] ++(0.5,0);
 \draw [dashed] (7.5,0) -- ++(1.5,0) coordinate(b);
 \draw (b) to [short, -] ++(0.5,0) coordinate(c)
 to [L, l=$L_{15}$] ++(0,4)
 (c) to [short, -] ++ (2,0)
 to [L, l=$L_{16}$] ++(0,4);
    \end{circuitikz}
    \caption{Schematic of the inductor array to generate the synchronisation signal. $L_1 = L_2 = ... = L_{24} = 82~mH$. The connection between the \modified{16} inductors connected in parallel and the voltage source is controlled by a square wave generated by a PIC18F47K42 microcontroller.}
    \label{fig:schematics}
\end{figure}
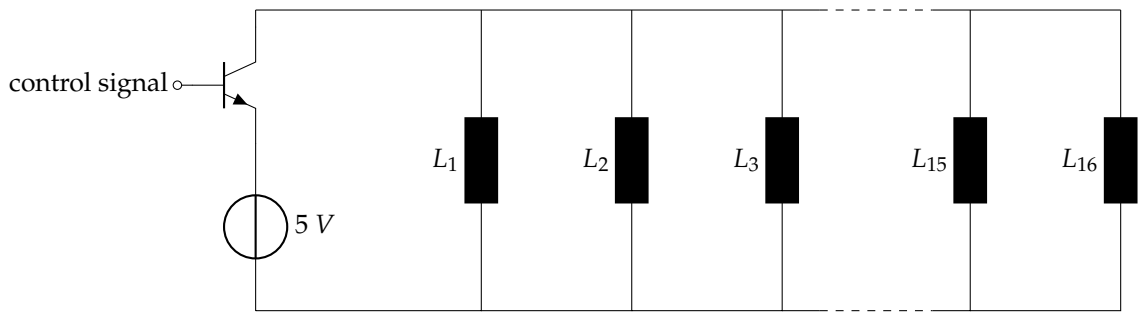

The induced magnetic field is generated by an array of identical inductors, connected in parallel. Each of the inductors generates the magnetic field for one IMU. We used inductors by Bourns (RLB0913-820K) with an inductance of 82~mH and a resistance of $212~\Omega$. These physical specifications result in a time constant $\tau$ of $390~\mu s$. \modified{The inductors used have an inductance manufacturing tolerance of 10~\%, which influences the  $\tau$ in the tenth of a $\mu s$ range.} The inductors are connected with a direct voltage source of 5 Volts. A transistor (MOSFET, maximum switching time 150~ns) makes or breaks the connection between the inductors and the power supply depending on the control signal. Our setup includes 16 inductors to be capable of synchronising the same number of IMUs (see figure~\ref{fig:schematics}). Theoretically, a larger number of inductors is also possible, provided a suitable voltage supply is selected. 

To control the transistor and thus the inductor circuit we use a $6~Hz$ square wave which is generated by a Microchip PIC18F47K42 microcontroller. \modified{The square wave signal ranges from 0 - 5 V and is generated by the microcontroller with a slew rate of 0.8~$\frac{V}{ns}$.} To ensure a consistent frequency we equipped this microcontroller with a GPS disciplined oscillator. This kind of oscillator has a very high stability. It is based on a temperature compensated crystal oscillator (TCXO) with a stability of 0.5~ppm. This TCXO is additionally corrected by the PPS signal of a GPS receiver, resulting in a stability of 1~ppb.

\begin{figure}[ht]
    \centering
    \includegraphics[width=14 cm]{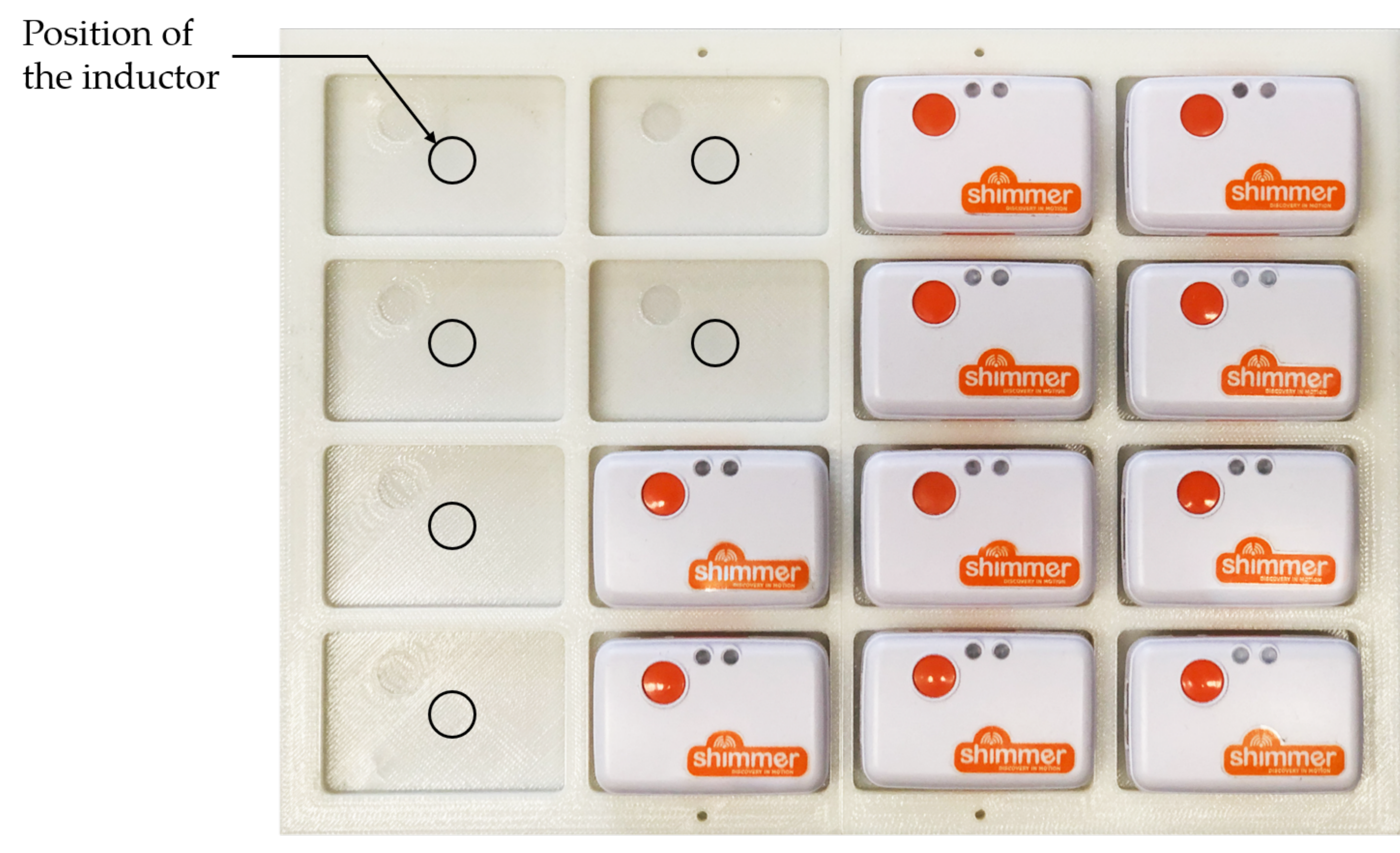}
    \caption{Docking station which ensures that each IMU is subjected to a comparable magnetic field.}
    \label{fig:dockingstation}
\end{figure}

The described circuit is placed beneath a 3D-printed docking station for the Shimmer3 IMUs, which has a precisely fitted notch for each IMU to ensure accurate positioning relative to the generated magnetic field. Beneath each notch an inductor is placed (see figure~\ref{fig:dockingstation}). Between the housing of an IMU and the associated inductor is one mm of plastic to subject the sensor to a magnetic field as powerful as possible. The positioning within the docking station assures that each IMU is subjected to a comparable magnetic field. Before and after each measurement, the IMUs are placed in this docking station and exposed to the synchronisation signal.

We implemented the described synchronisation algorithm in Python 3.7. After a measurement the data of each IMU is processed and synchronised by this algorithm using the explained procedure.

\subsection{Validation}
\label{section:val}
To validate the proposed technique, a suitable setup was designed. We used the described setup (see section \ref{section:tech}) and controlled the array of inductors with a square wave with a frequency of 6~Hz. Additionally, the same signal is applied to the analog-to-digital converter (ADC) channel of a Shimmer3 IMU as a reference signal. This way the IMU is able to measure the magnetic flux density and the signal controlling the inductor circuit simultaneously.

The Shimmer3 IMUs contain a 16-bit magnetometer LSM303AHTR by STMicro, which has a maximum sampling rate of 100~Hz and measures the magnetic flux density in a range of $\pm 49.152$~Gauss. They are also equipped with a \modified{12-bit ADC (MSP430 by Texas Instruments)} input channel, which allows them to measure an applied voltage synchronously with inertial metrics. All metrics of one unit are recorded relative to the same local time. As the maximum sample rate of the ADC channel is limited when transmitting the data via Bluetooth, we logged the data on a SD card inside the device. For this experiment we used a sampling rate of 1310~Hz for the ADC. We are therefore able to determine the actual time of the first edge of the control signal with a maximum uncertainty of $ \frac{1}{1310}=763 \mu s$. 

Based on the described setup we design three experiments. The objective of the first experiment is to test the accuracy of the proposed technique. To achieve this three different Shimmer3 units are used. The units are subjected to the magnetic and the control signal for a time period of ten seconds. This procedure is repeated 200 times. Afterwards, the starting point of each procedure is determined in two ways: the start time $t_{calc}$ is calculated based on the magnetometer values with our described novel approach. As a reference, the ADC values are used to determine the start time $t_{meas}$ of the first edge of the control signal. Based on the acquired measures and the results of the algorithm we calculate various parameters to assess the accuracy:

\begin{itemize}[leftmargin=*,labelsep=5.8mm]
    \item the difference $\Delta t$ between the calculated and the measured start time of the first edge of the square wave:
        \begin{equation}
           \Delta t= t_{meas} - t_{calc}
        \end{equation}
    \item the mean $\mu_{\Delta t}$ and standard deviation $\sigma_{\Delta t}$ of all differences $\Delta t$ for one IMU
    \item the number of hits $k $ for one synchronisation process
    \item the mean $\mu_{k}$ and standard deviation $\sigma_{k}$ of $k$ of all tests of one IMU
    \item the coefficient of determination $R^2$ of the first order polynomial fitting 
    \item the $\mu_{R^2}$ and standard deviation $\sigma_{R^2}$ of $R^2 $ of all tests of one IMU
\end{itemize}

A second experiment is designed to evaluate how the number of measured $hits$ influences the accuracy of the algorithm and how $R^2$ changes with an increasing number of $hits$. The duration of the synchronisation process is varied, ranging from one to 30 seconds, in steps of one second. This leads to an increasing number of detectable $hits$, as this quantity is linearly dependent on the process duration. The data is collected using the same setup and circuit as in the first experiment but with one IMU. For each individual duration, the experiment is performed ten times. To analyse the results, we use the previously introduced measures $\Delta t$ and $R^2$.

With the third experiment we want to examine the drift of the local IMU clocks. The purpose of this examination is to evaluate whether our proposed linear drift compensation is reasonable. Eight IMUs are exposed to the described synchronisation signal in regular intervals. The synchronisation process is triggered every 300 seconds over the course of one hour. The triggering is performed by a python script running on a Windows 10 PC. Before each trigger we save the current system time with microsecond precision as a reference. After the measurement, the start times of the synchronisations are determined in the local time contexts of the individual sensors. We use the deviations between these local IMU timestamps and the saved reference computer clock timestamps to analyse the temporal drift of the local clocks. 
To ensure that this system can perform accurate measurements we have taken several measures:
We adjusted the settings of the 'windows time service' according to an official support document \cite{Dahavey.17.03.2021} to increase the accuracy of the PC's system clock.
The delay between the start of the synchronisation procedure and the execution of the corresponding python trigger function is kept constant with an USB TTL Module by Black Box ToolKit Limited (Sheffield, United Kingdom). This device ensures a constant delay of below 1~ms between the call of a software method and the setting of a 5V level on an output line. The control circuit for the synchronisation procedure is set to start the procedure when the USB TTL Module signal is received, which results in a constant delay. \modified{The used IMUs are all placed next to each other in the same room, so the ambient temperature of each sensor unit is the same.}


\section{Results}
\label{section:3}



The purpose of experiment one was to evaluate the accuracy of the proposed algorithm. Table~\ref{table:res1} presents the descriptive statistics of the acquired metrics which were defined in section \ref{section:val}. The first four columns show the measured offset $\Delta t$. Surprisingly, all three sensors show a significant offset. The mean offset is negative, which means the estimated start happens after the measured start of the sequence. Furthermore, the low standard deviation indicates that the offset of the procedure is constant. To be able to define a confidence interval, we tested the acquired offsets for normal distribution using the Shapiro-Wilk test ($\alpha = 0.05$). The results of sensor one (p=0.114), sensor two (p=0.636) and sensor three (p=0.09) indicate a normal distribution of the offsets. 

Turning now to the evaluation of the number of detected $hits$ per synchronisation procedure, which is shown in column five and six of table~\ref{table:res1}. Each IMU measures a comparable average amount of $hits$ with small relative deviations. As each procedure lasts ten seconds, the IMUs measure a $hit$ every 387~ms on average.

The goodness of fit of the performed polynomial fittings can be analysed with the metrics in the columns seven and eight of table~\ref{table:res1}.
On average all fittings show an almost perfect linear relationship with a negligible deviation. These results prove that the algorithm and the electronic setup worked as intended, because the $hits$ were reliable produced and detected.

\begin{table}[h]
    \caption{Results of experiment 1: Descriptive statistic for the measured offset $\Delta t$, the number of $hits$ $k$ and the coefficient of determination $R^2$}
    \label{table:res1}
    \centering
    \begin{tabular}{cccccccc}
    \toprule
    \textbf{$\mu_{\Delta t}$ [ms]}	& \textbf{$\sigma_{\Delta t}$ [ms]}	& \textbf{Min $\Delta t$ [ms]} &
    \textbf{Max $\Delta t$ [ms]} & \textbf{$\mu_{k}$}  & \textbf{$\sigma_{k}$}& \textbf{ $\mu_{R^2}$}  & \textbf{$\sigma_{R^2}$}\\
    \midrule
    -2.069  & 0.302     & -1.45     & -3.15     & 26.33     & 1.619 & 0.999 & 6.702 e-08\\
    -2.073  & 0.282     & -1.31     & -2.92     & 25.762    & 1.503 & 0.999 & 7.862 e-08\\
    -2.07   & 0.287     & -1.37     & -2.37     & 25.546    & 1.919 & 0.999 & 3.478 e-08\\
    \bottomrule
    \end{tabular}
\end{table}

\begin{figure}[h]
    \centering
    \includegraphics[width=15 cm]{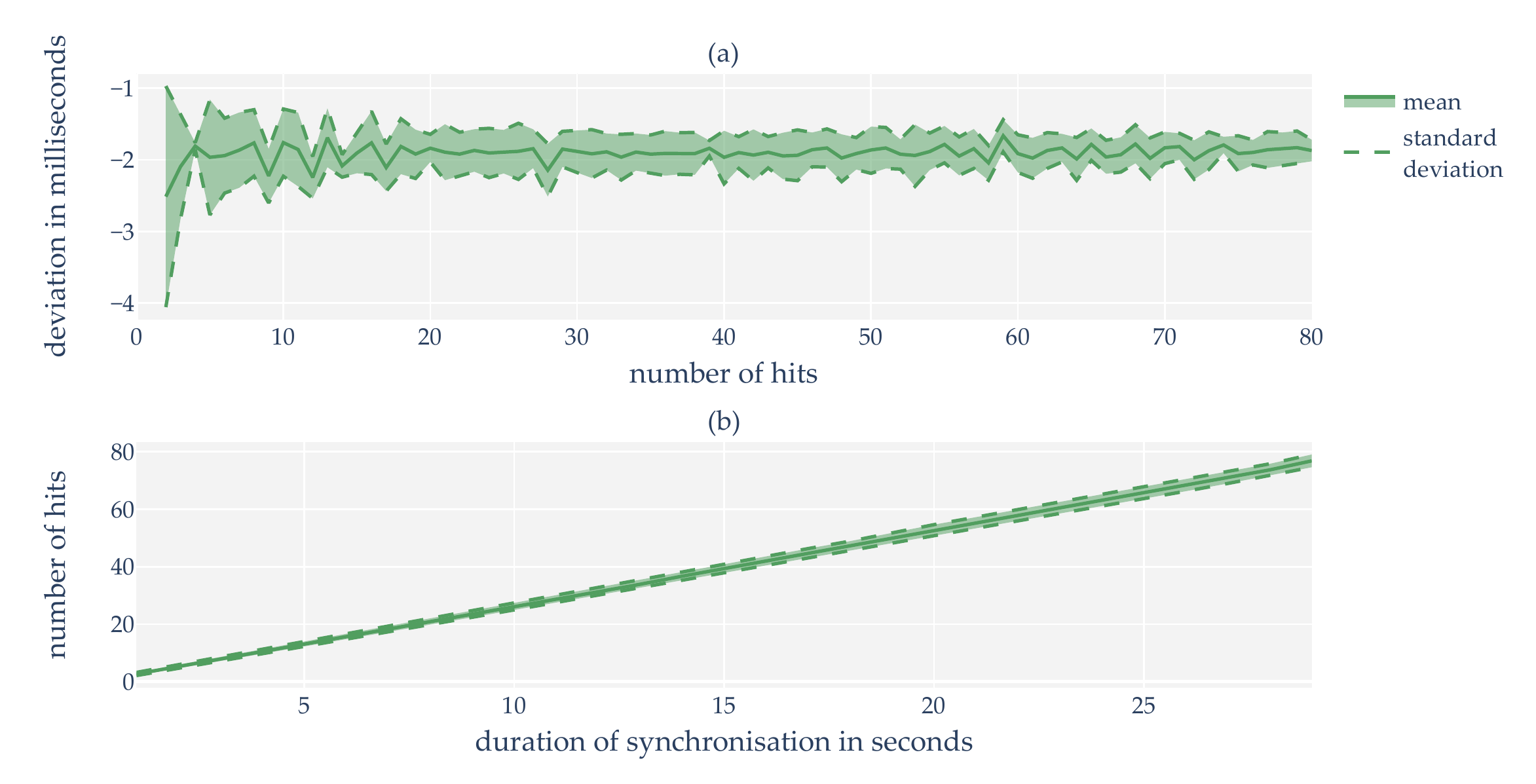}
    \caption{Results of experiment 2. (a) Calculated offset $\Delta t$ dependent of the number of $hits$. (b) Mean and standard deviation of the number of $hits$ dependent of the duration of the synchronisation procedure. }
    \label{fig:results2}
\end{figure}

The second experiment aims to analyse the influence of the number of acquired $hits$ on the accuracy of the algorithm. Figure~\ref{fig:results2}~(a) displays the scatter diagram of the relationship between the number of $hits$ of a synchronisation procedure and the achieved accuracy. The mean accuracy fluctuates between three and 20 $hits$ but stays consistent for a higher number of $hits$. For more than 20 $hits$ there is no notable increase in accuracy. In addition, figure~\ref{fig:results2}~(b) shows the interaction between the duration of the synchronisation process and the number of detected hits. The relationship between the two variables can be clearly described as linear. 

Additionally, the connection between achieved goodness of fit and number of available $hits$ was investigated. For all performed measurements, $R^2$ was greater than 0.999. It is apparent that the goodness of fit shows no considerable changes due to an increased number of $hits$.

\begin{figure}[h]
    \centering
    \includegraphics[width=15 cm]{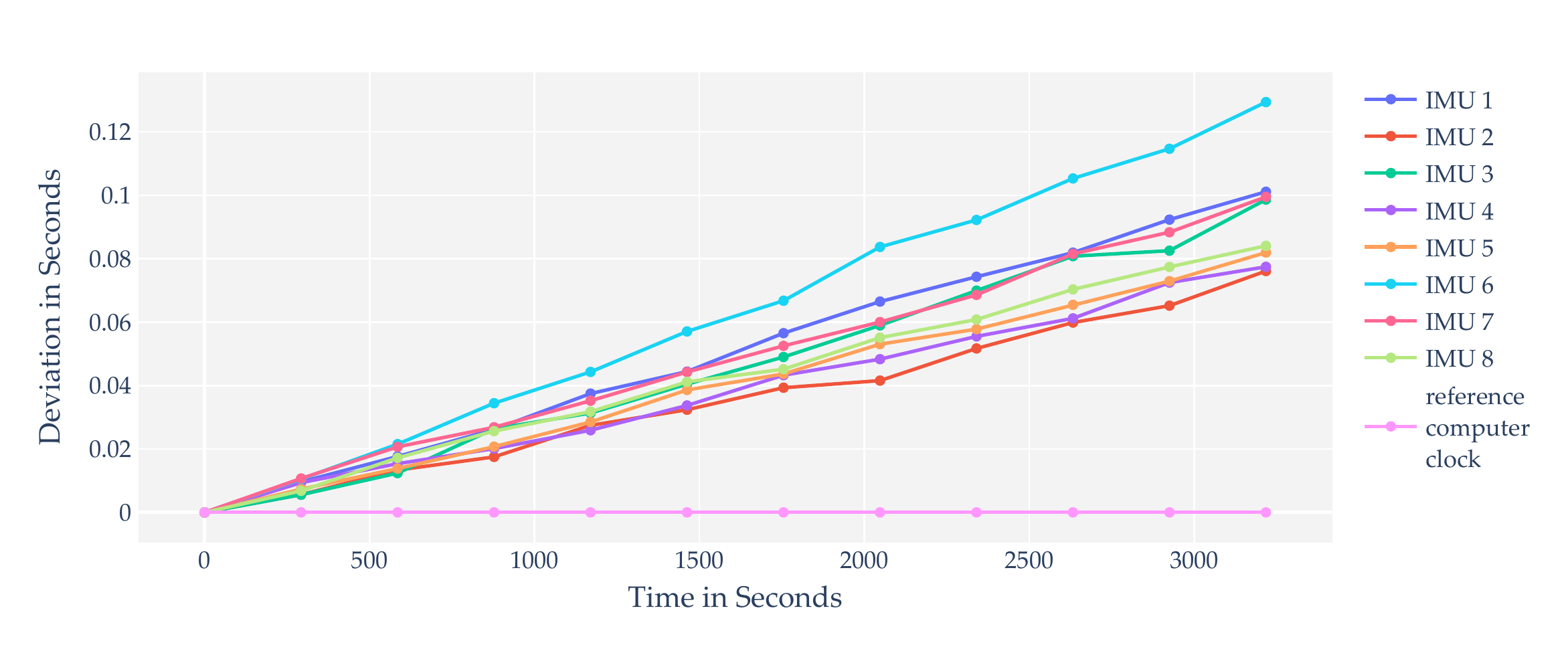}
    \caption{Results of experiment 3. Deviation between the local IMU clocks and the reference computer clock over the course of one hour.}
    \label{fig:results3}
\end{figure}

The intention of the third experiment was to evaluate the temporal drift behaviour of the local IMU clocks. Figure~\ref{fig:results3} shows the deviation between the IMU clocks and the reference computer clock for each of the twelve performed synchronisation procedures. After 3600 seconds most of the local clocks are 76 to 100~ms ahead of the reference clock with one outlier which is 130~ms ahead. We performed a least squares regression line fitting with the deviations of each IMU. The fitting's $R^2$ values cover a range from $R^2 = 0.993$ to $R^2=0.998$, which indicate a high degree of linear relationship.

\section{Discussion}
\label{section:4}

The used experimental setup allows us to record suitable measurement values in a repeatable manner. The docking station for the IMUs ensures a similar distance between the individual IMUs and the inductors, which leads to a comparable magnetic field at the sensors. The inductance of the inductors and the measurement accuracy of the sensors are appropriate to measure sufficiently precise values at the defined distance. Furthermore, the magnitude of the inductance is adequate to minimise the influence of regular environmental influences on the generated field. One can further improve the measurement setup by using inductors with a narrower manufacturing tolerance.

The results of the first two experiments allow several interesting conclusions. An important finding is the low deviation of the number of hits $k$ during all experiments with comparable duration. This indicates a reliable $hit$ detection of the used algorithm. Additionally, these findings suggest that the used electronic setup produces a predictable amount of $hits$ over a given time period. This consistency is achieved by the explained relation between sample rate of the IMUs and the sample rate of the square wave. Concluding, we can say that the proposed approach is reliable.

Related to that, the second experiment shows that there is no further improvement in accuracy for a number of $hits$ larger than $k > 20$. With the proposed setup and a control signal frequency of 6~Hz, a minimal synchronisation duration of around eight seconds can be recommended. 

The most important result is the measured accuracy. Surprisingly, there is a mean offset of about -2~ms, with low scattering. Several factors might explain this observation. Firstly, we excluded the possibility of an implementation issue by testing the algorithm with synthetic data, which resulted in a mean accuracy noticeable below one millisecond. Secondly, we analysed if the design of the electric circuit significantly decelerates the signal transmission, which was shown also not to be the case. 
Thirdly there might be a problem in the used measurement setup. We used the Shimmer3 IMUs to measure the control signal of the circuit and the magnetic field itself. This setup was chosen on the assumption that the measurements of the individual sensors inside the Shimmer3 happen synchronous. But these channels have a large difference in acquisition speed. The sampling rate of the analog input channel is fifteen times higher compared to the rate of the magnetometer. Therefore, for each magnetometer reading there are 15 voltage readings. Depending on the order in which different events are processed in the IMU's microcontroller there could emerge an offset between measuring and reading the acquired magnetic flux density. For example, the delayed reading might be assigned to the timestamp of the reading and not to the timestamp of the measurement. This hypothesis is supported by the fact, that all three used Shimmer IMUs had a comparable mean offset with a small standard deviation. Hence it appears, the Shimmer3's program sequence causes a constant offset in the measurement setup. As this offset is of deterministic nature and constant, it can be directly compensated.

To calculate the possible synchronisation accuracy of the proposed approach we have to consider the following aspects. It is important to notice that the calculated offsets are normally distributed. As a result, it is highly unlikely that an offset occurs differing more than three times the standard deviation from the mean value. Also, we have to mention the inaccuracy of the used reference. The starting time of a control signal's edge is measured with 1310~Hz, which results in a maximum sampling uncertainty of one sample, which corresponds to 0.763~ms. Taking these observations into account and subtracting the constant offset, the accuracy of our algorithm is below 2.6~ms. This is substantially lower than the accuracy of a normal event-based synchronisation, which is 10~ms for sensors sampling at 100~Hz. Therefore, the presented approach provides a significant improvement over conventional methods. It features improved accuracy, a precise event generation and precise event detection, combined with a fully automatic time synchronisation algorithm.

The third experiment showed that the temporal drift of the IMU clocks can be described as linear, \modified{if the ambient temperature is constant. However, as already mentioned in section~\ref{section:1}, there are several influences that affect the frequency of an oscillator and thus the timing of a sensor. For oscillators whose components are not kept at a constant temperature by additional electronics, the environmental factor with the greatest influence is temperature \cite{Tirado-Andres.2019}. The resulting frequency change depends on the current temperature, therefore the influence of a temperature change on the time measurement cannot be described linearly. This behaviour is comparable for identical oscillators, but it scatters in a defined range. In general, a temperature change during a measurement can be caused by two influences: Either by the waste heat of the electronic components or by a change in the ambient temperature. The approach described cannot compensate for this non-linear component of the temporal drift on the basis of two synchronisation events. However, additional factors must be taken into account, which reduce the influence of a temperature change in the defined application: Identical sensors are used, which means that the influence of the temperature is comparable within a certain tolerance range. The body temperature of a test person is likely to differ from the ambient temperature and thus influence the sensors. However, the sensors are not applied directly to the skin and are also insulated by a mount and housing. Thus, a change in body temperature does not directly result in a change in temperature at the sensor. This problem can be compensated for by attaching the sensors to the test person a certain time before the measurement. In this way, the sensors are already exposed to a corresponding temperature before the measurement. During the measurement, only minor temperature changes occur. In addition, the measurement should be carried out in an environment with a constant ambient temperature. During this acclimatization phase before the measurement, the sensors should also be switched on already and measure. As a result, the electronic components will no longer experience any significant temperature changes. These additional steps should allow the proposed approach to synchronise the data with sufficient accuracy. However, further investigations should be carried out to determine the magnitude of the temperature changes in the given application.} 

\modified{Based on the results of the third experiment, }it is possible to compensate the \modified{linear component of the} temporal drift as proposed by performing a synchronisation procedure before and after a measurement. Without a reference clock, it is possible to compensate the \modified{linear component of the} temporal drift of the IMUs relative towards each other by choosing an arbitrary reference among the IMUs. For a measurement using just IMUs this is a suitable approach, as it greatly reduces the temporal error between the IMUs. If one wants to perform a measurement with additional systems, it is necessary to include a reference clock into the system.

\modified{Since the sensors must be stored in the docking station (see figure~\ref{fig:dockingstation}) for a synchronisation event, usually only two synchronisation events can be applied, as additional events would be disruptive for the measurement process. Therefore, the presented approach has the same disadvantages as any other synchronisation approach based on only two events. For the proposed use case of human motion analysis, the advantages outweigh the disadvantages. Especially the influence of the ambient temperature on the oscillators is likely to cause considerable deviations, due to the natural fluctuations of the outer temperature for measurement duration of multiple hours. Since human motion analysis tasks normally cover a period of several hours at most, we can assume that this will not raise a problem. The chosen synchronisation approach allows correction of timestamps only after the second synchronisation event has been acquired. Accordingly, synchronisation during the measurement is not possible. Since the analysis of the data is usually performed after the measurement, this is not a problem for the defined use case.}

Since the presented approach, as explained in section \ref{section:2}, is based on the asymptotic course of the magnetic flux density during an inductor's transient response, it can also be applied to other IMUs from different manufacturers, as far as the IMU meets the following requirements: It has to be equipped with a magnetometer with a measuring range and resolution that resolves the magnetic field changes sufficiently. In addition, the sample rate of the magnetometer has to be significantly higher compared to the frequency of the inductor circuit's control signal to be able to map the signal accurately. We recommend testing IMUs from other manufacturers for delays in their internal measurement process to be able to identify and compensate for offsets comparable to those of the Shimmer3 IMUs. That said, our approach can also be used for sensors without a magnetometer that feature an ADC. The measurable voltage on a capacitor while it is being charged or discharged can be described by the same formula as the transient response of an inductor. Therefore, one could synchronise sensors with an ADC with sub-sample accuracy.

\section{Conclusions}
\label{section:5}

In this work, we propose a novel event-based synchronisation procedure for IMUs, \modified{suitable for human motion analysis tasks}. Such a process has to be able to reach sub-sample accuracy. We developed a suitable approach based on a common magnetic field, which the individual IMUs are exposed to. To evaluate this approach, we tested different measures like accuracy, reliability and the relationship between accuracy and required duration of the synchronisation procedure. The results show that the process is capable of reliably generating and detecting synchronisation events. The accuracy was also evaluated in relation to the duration of the process and it was found that a duration of more than eight seconds results in no measurable improvements. \modified{The temporal drift of the sensors could be confirmed as linear at a constant ambient temperature. However, it should be additionally investigated which temperature variations the sensors experience during a motion analysis, in order to be able to consider further non-linear influences on the temporal drift.} Finally, taking into account the temporal resolution of the used reference, an offset of less than 2.6 ms could be proven. A constant offset was found, probably due to the firmware of the used Shimmer3 IMU. In further investigations it would therefore be useful to evaluate the presented approach with sensors from different manufacturers in order to ensure a comparable effect.

 \vspace{6pt} 

\authorcontributions{conceptualization, A.S. and M.M.; methodology, A.S. and M.M.; software, A.S.; validation, A.S. and M.M; formal analysis, A.S. and M.M.; investigation, A.S.; resources, M.M; data curation, A.S.; writing--original draft preparation, A.S.; writing--review and editing, M.M.; visualization, A.S.; supervision, M.M.; project administration, M.M.; funding acquisition, M.M.}

\funding {: The article processing charge was funded by the Baden-Württemberg Ministry of Science, Research and Culture and the Ulm University of Applied Sciences via the funding program Open Access Publishing}

\acknowledgments{In this section you can acknowledge any support given which is not covered by the author contribution or funding sections. This may include administrative and technical support, or donations in kind (e.g., materials used for experiments).}

\conflictsofinterest{The authors declare no conflict of interest.} 

\externalbibliography{yes}
\bibliography{references.bib}

\end{document}